 \documentclass[12pt,draftclsnofoot,onecolumn]{IEEEtran}

\usepackage{amsmath}
\usepackage{algorithm, amsbsy,amsmath,amssymb,epsfig,bbm,mathrsfs, bbm} 
\usepackage{amsthm}
\usepackage{verbatim}
\usepackage[noadjust]{cite}
\usepackage[subfigure]{graphfig} 
\usepackage{amsmath,mathtools}

\usepackage{algorithm}
\usepackage{algpseudocode}
\usepackage{amsmath}
\usepackage{graphics}
\usepackage{epsfig}
\graphicspath{{./figures/}}
\theoremstyle{definition}

 \usepackage{epstopdf}
\usepackage{geometry}
\usepackage{xcolor}
\allowdisplaybreaks
\usepackage{xcolor}
\allowdisplaybreaks

\usepackage{color}
\definecolor{MyDarkBlue}{rgb}{0,0.08,0.45}
\definecolor{yellow}{rgb}{0.99,0.99,0.70}
\definecolor{myback}{RGB}{204,232,207}  
\definecolor{white}{rgb}{1.0,1.0,1.0}                             
\definecolor{black}{rgb}{0.00,0.00,0.00}

\begin{document}
 
\title{Intelligent Reflecting Surface (IRS)-Enabled Covert Communications in Wireless Networks}
   \author{Xiao Lu, Ekram Hossain, Tania Shafique,  Shaohan Feng, Hai Jiang, \\ and Dusit Niyato 
 }   	\vspace{-8mm} 
   	\vspace{-8mm} 
\maketitle

\vspace{-15mm}
\begin{abstract}

With growing security threats to the evolving wireless systems, protecting user privacy becomes progressively challenging. Even if the transmitted information is encrypted and the potential wiretap channel is physically limited, the raw data itself, such as transmitter position and transmission pattern, could expose confidential information. In this context, covert communication that intends to hide the existence of transmission from an observant adversary emerges as a practical solution. However, existing covertness techniques ineluctably consume additional resources such as bandwidth and energy, which burdens system deployment. In view of this concern, we propose an intelligent reflecting surface (IRS)-based approach to enhance communication covertness. The core idea is making use of a smartly controlled metasurface to reshape undesirable propagation conditions which could divulge secret messages.  To facilitate the understanding of the proposed idea, we first provide an overview of the state-of-the-art covert communication techniques. Then, we introduce the fundamentals of IRS and elaborate on how an IRS can be integrated to benefit communication covertness. We also demonstrate a case study of the joint configuration of the IRS and the legitimate transmitter, which is of pivotal importance in designing an IRS-enhanced covert communication system. Finally, we shed light on some open research directions.

\end{abstract}
\begin{IEEEkeywords} \vspace{-2mm} 
Wireless security and privacy, physical-layer  security, covert communications, reconfigurable metasurface, interference cancellation
\end{IEEEkeywords}
\vspace{-2mm}

\section*{Introductions}
Provisioning secured communication becomes unprecedentedly challenging owing to the threat of technology integration. 
The ubiquitousness of access interfaces and utilization of shared spectrum in an open wireless medium make the soaring volume confidential information (e.g. financial account, identity authentication, and business secret) 
 more exposed to malicious attackers, the goal of which is to intercept sensitive and private data.  Therefore, ensuring the reliability and security of wireless data remains one of the most important tasks in developing future generation networks and has drawn increasing attention from wireless 
communities~\cite{I.2018Ahmad}.
 
The current practice of wireless security mainly relies on application/transport-layer encryption. However, securing wireless communications with encryption  
faces the following challenges: 1) the standardized protocols adopted for public networks make a large number of entities confront the same threat; 2) the security level of encryption protocols could be compromised if eavesdroppers have powerful computational capacities as decryption involves solving mathematical problems; and 3) distribution and management of cryptographic keys are difficult in decentralized networks with random access and mobility. 


To cope with these difficulties of encryption, physical-layer security (PLS) approaches have drawn significant research attention over the past years.  
Essentially, PLS approaches safeguard information by only exploiting the fundamental nature of the wireless medium (i.e. interference, noise, and fading), which avoids extra signaling and communication overheads incurred by  encryption in the higher layer(s).
There are two remarkable research tendencies on PLS, namely, {\em information-theoretic secrecy} (ITS) \cite{Y.2009Liang} and  
{\em covert communication}~\cite{A.Sep.2013Bash}.
ITS approaches aim to achieve a positive secrecy rate (i.e. the rate difference of a legitimate channel and an eavesdropping channel), at which information can be conveyed confidentially. Nevertheless, merely preventing transmission from being deciphered is not sufficient from the perspective of  privacy protection. 
There appear progressively more circumstances where revealing the position, movement, or even the existence of communication is crippling or even fatal. For example, exposure of business activities could bare commercial secrets. 
This raises the need for covert communication, also known as \emph{low probability of detection (LPD) communication} or \emph{undetectable communication}, the objective of which is to shelter the presence of a legitimate transmission from a vigilant adversary while maintaining a certain covert rate at the intended user\footnote{In this article, covert communication refers to physical-layer techniques that hide wireless transmission over covert channels. This is different from the concept of {\em covert information} techniques that conceal a secret message in a cover medium (e.g. text, image and audio/video message) instead of masking the transmission behavior.}.

\vspace{-2mm}
\section*{Overview of Covert Communication}


Notably, covert communication offers three major advantages as follows: First, covertness techniques guarantee a stronger security level compared to ITS. If a communication link is hidden from an adversary, the information carried is immune from interception. Secondly, in contrast to encryption, the performance of covert communication does not rely on the adversary's competence. In other words, the achievable security level will not be degraded even if the adversary has powerful information processing capability. 
Thirdly, covertness techniques can be implemented either as  alternative or complementary solutions for upper-layer security and privacy techniques, such as steganography and encryption.  
This section first introduces the principles of covert communication and then presents an overview of the existing techniques.



  
 \vspace{-2mm} 
\subsection*{Understanding Covert Communication}

Consider a general point-to-point communication scenario where a legitimate transmitter (Alice) intends to deliver a message wirelessly to the target receiver (Bob) without being detected by an adversary (warden Willie). Willie monitors the wireless channel with the aim to detect whether Alice is on transmission or not. Hence, Willie faces a binary decision between null hypothesis $\mathcal{H}_{0}$ that Alice is mute and the alternative hypothesis $\mathcal{H}_{1}$ that Alice is transmitting. 
For such a purpose, Willie can perform statistical hypothesis testing based on 
the average power received in a time slot denoted  as $\bar{P}_{W}$. $\bar{P}_{W}$ contains the received interference power $I_{W}$ and noise power $\sigma^2_{W}$ in the case of $\mathcal{H}_{0}$ and additionally contains the received signal power $S_{W}$ from Alice in the case of $\mathcal{H}_{1}$. Let $\mathcal{D}_{0}$ and $\mathcal{D}_{1}$ denote the decisions of Willie in favor of $\mathcal{H}_{0}$ and $\mathcal{H}_{1}$, respectively. The decision of Willie follows
 a threshold-based rule which advocates $\mathcal{D}_{0}$ and $\mathcal{D}_{1}$ when $\bar{P}_{W}$ is smaller or greater than a predefined threshold  $\tau$, respectively.
 According to this rule, erroneous decision occurs in two circumstances: 1) Willie sides with $\mathcal{D}_{1}$ when $\mathcal{H}_{0}$ is true, i.e. {\em false alarm}, and 2) Willie sides with $\mathcal{D}_{0}$ when $\mathcal{H}_{1}$ is true, i.e. {\em  mis-detection}. 
 The total probability that Willie makes erroneous decisions (i.e. including false alarm and mis-detection) can be interpreted as the {\em covert probability} for transmissions from Alice to Bob. It is evident that the covert probability is influenced by the uncertainties
of $S_{W}$, $I_{W}$, and $\sigma^2_{W}$.

Fig. 1 illustrates the impacts of different parameters. The blue and orange lines represent the probability density function (PDF) of  $I_{W}+\sigma^2_{W}$ 
with smaller variance and larger variance, respectively. 
For each case, the mis-detection probability and false alarm probability 
can be represented by the left and right shadow areas, respectively. We can observe that, for a given $\tau$, it is possible to increase the 
mis-detection probability by decreasing $S_{W}$, and increase both  the mis-detection probability and false alarm probability 
by enlarging the variance of interference and noise. Thus, communication can be carried out more covertly with reduced signal leakage to Willie and/or with larger interference plus noise fluctuations.

\begin{figure} \label{PDF}
\centering
\includegraphics[width=0.5\textwidth]{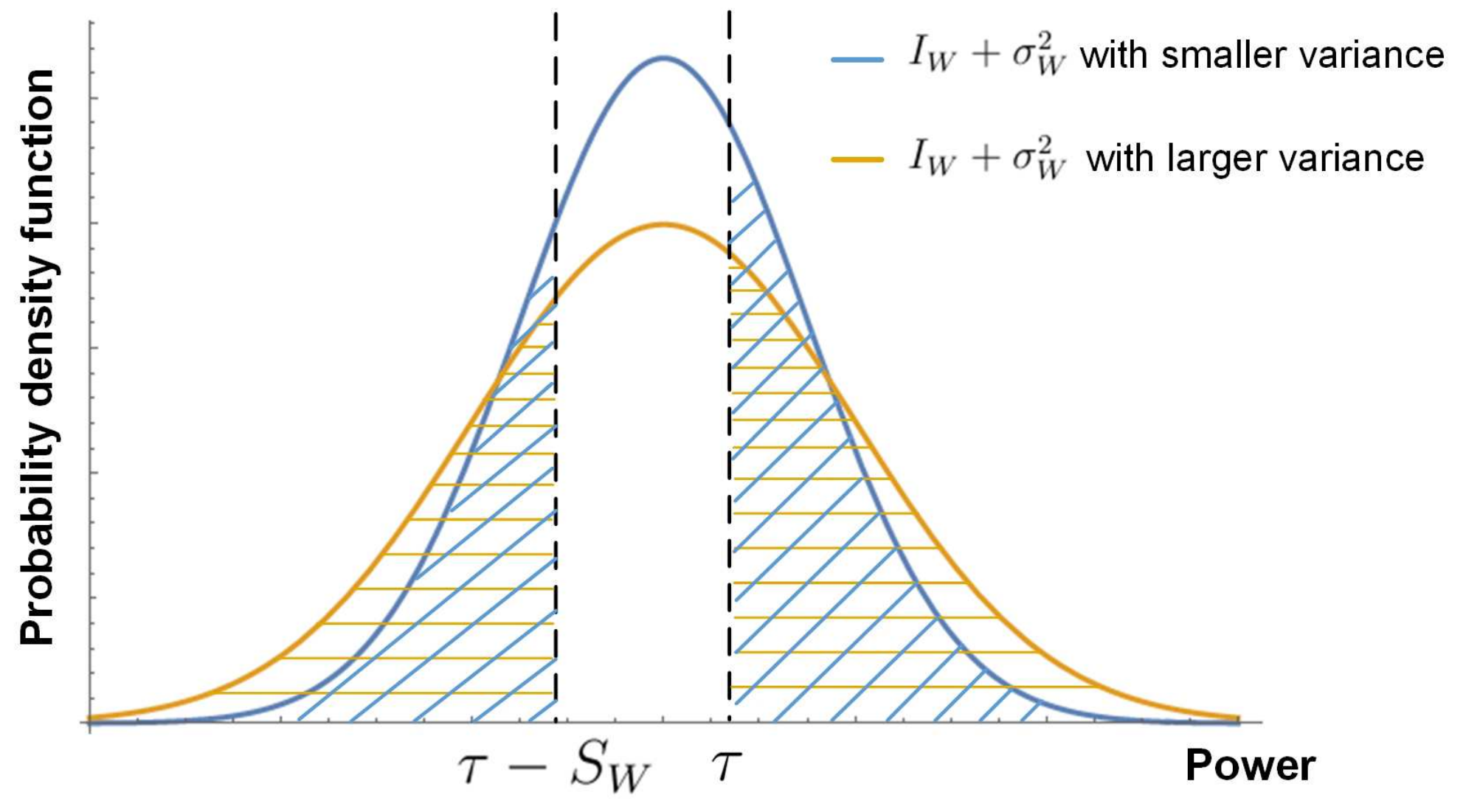} \vspace{-5mm}
\caption{Illustration of the impacts of system parameters for covert communications.}  \vspace{-8mm}
\end{figure}

\vspace{-3mm}
\subsection*{Overview of Covertness Techniques}

By exploiting the properties of covert communication, different approaches have been developed to enhance covertness performance, which are reviewed below.

\subsubsection*{Multiple Antennas}

Taking advantage of spatial degrees of freedom,  multiple-antenna techniques can be utilized to improve the stealthiness of wireless channels 
through directional transmission~\cite{X.2019Zheng}.  
This can be realized by means of beamforming
to produce spatial selectivity. 
In particular, a beamformer adjusts the relative phase and amplitude of the signals on each element of an antenna array such that the superposed radiation pattern is constructive in the desired direction and destructive in the other directions. 
As a consequence, the transmitted signals can be concentrated towards the desired recipient to enhance the achievable rate and concurrently nulled at the adversary for LPD. Multiple-antenna techniques have high deployment scalability as it is a transmitter-side implementation.
Beamforming performance is largely dependent on the availability of instantaneous channel state information (CSI). 
The inaccuracy of CSI at a multi-antenna transmitter due to estimation error could result in a high probability of signal leakage to the adversary and thus degrade the covertness performance.     
However, the negative impact of imperfect CSI  can be mitigated as the number of transmit antennas becomes massive, e.g. more than one hundred~\cite{E.Bjornson}. 
The highly correlated channels in massive antenna region  
render minimal CSI estimation errors and hence high beamforming
resolution.  
Channel hardening effect that makes effective channel gains deterministic is an additional attribute of massive antennas that can be exploited to provide reliable covert rate. 

\subsubsection*{AN Generation}
 
Random AN can be generated to increase interference dynamics, 
deliberately 
misleading the decisions of the adversary regarding the existence of any covert transmission. The key to a successful AN design is to avert 
the negative impact of jamming signals on legitimate channels \cite{R.2018Soltani}. 
For this, multiple-antenna techniques can be exploited to produce AN nulling in the directions of legitimate users. A more robust covert performance can be achieved if the position of the adversary is known so that the detectivity of the adversary can be corrupted to the largest extent through directional jamming.

An outstanding feature of the AN generation approach is its flexible implementability. In practice, AN can be generated by different entities. Some common ways of realizing AN generation are described as follows.
  
\begin{itemize}

\item 
{\em Cooperative jamming} employs a third-party device (e.g. power beacon and drone radio transmitter) that functions as a helper to jam adversary's channel.  
One or more friendly jammers can coordinate with Alice to disturb Willie's channel while causing minimal impact on the legitimate transmission. 
Cooperative jamming incurs synchronization and communication overhead for transmit power control. Moreover, the use of cooperative jammer(s) sacrifices deployment scalability and may not work efficiently in the presence of mobility. 

\item 
{\em Full-duplex jamming} can support concurrent information reception and in-band AN generation with a full-duplex receiver. 
This approach surmounts the control overhead and mobility 
issues of cooperative jamming at the cost of loopback self-interference from transmit to receive RF chains.  
Thanks to the recent advance of full-duplex techniques in multiple domains (e.g. antenna interface, analog baseband, and digital processing), self-interference can be suppressed to a tolerable extent with a viable expense.  

\item 
{\em AN injection} is a sender-side technique capable of transmitting information signals and AN simultaneously.  
Ideally, AN is constructed to be orthogonal to the legitimate channel such that only Willie's channel is affected. 
The crux of AN injection is to balance the trade-off between covertness and information rate by optimizing the transmit powers of jamming and information signals.

\end{itemize}

In addition, hybrid approaches can be explored for performance enhancement. 

\subsubsection*{Cooperative Relaying}
Cooperative relaying relies on cooperation from intermediate  node(s) to facilitate undetectable communication.  
For legitimate communication,
the access distance has a profound effect on covertness.  
For long-distance transmission, high transmit power is required to attain a target rate that unavoidably compromises the covertness.
Cooperative relaying remedies this issue by  multi-hop forwarding. The rationale is to shorten the access distance of each hop 
so as to keep the required transmit power low, rendering a low detection probability by Willie. As cooperative relaying-based covert communication involves the use of the third-party device(s) as the relay(s), its deployment scalability is relatively low.
 

\subsubsection*{Spread Spectrum}

The spread spectrum approach facilitates covertness by  suppressing the average power spectral density (PSD) of the transmitted signal below the noise floor level.
Specifically, the information is modulated on a sequential noise-like wave, namely pseudo-noise sequence, which considerably spreads the transmission bandwidth compared to the one required by normal narrowband signals.
As a result, it is difficult for an adversary to discriminate the information-bearing signals from noise, which significantly lowers the signal detectability.
Typical modulation techniques adopted for bandwidth spreading include {\em direct sequence} which spreads the transmitted signal over multiple frequency channels and {\em frequency hopping} which randomly and speedily switches the transmission channel across a fairly wide frequency range.  
Generally, direct sequence is more immune to malicious detection as the PSD of the transmitted signal is continuously kept low. Frequency hopping is more exposed as it makes use of narrow-banded signals with high PSD on any frequency hop.  
In addition to LPD, the frequency diversity empowered by spread-spectrum signal offers robustness of covert communication against fading.
The spread spectrum approach has high deployability as it is a sender-side manipulation.
 
\subsubsection*{Millimeter-Wave Communications}

Operating at the frequency bands between 30-300 Gigahertz (GHz), millimeter-wave (mmWave) communication features steerable  
narrow beam,  i.e. precise angular resolution can be realized by moderately small antenna dimensions. 
The directionality of the narrow beam naturally  benefits covertness as signal leakage due to imperfect beam patterns towards the off-boresight
directions can be suppressed. To intercept mmWave communication, an adversary can only detect the misaligned beam, which exhibits an on and off behavior where the bursty beam arrives intermittently \cite{G.2014Andrews}. 
This distinguishing beam pattern effectively disrupts the detectability of an adversary. Furthermore, the ultra-wide bandwidth of mmWave compared to microwave allows high flexibility in the frequency range of legitimate transmission. Scanning signals on a wide spectral ambit imposes 
a great amount of overhead for signal detection at an adversary.  
The downside of short wavelength comes to weakened scattering and diffraction abilities which make mmWave attenuate acutely and susceptible to obstacles. Moreover, the Doppler shift of mmWave is strong even at walking speed. Hence, 
mmWave-based covert communication has low deployment stability as
the covert rate of mmWave communication is vastly affected by the availability of line-of-sight channels and mobility.
 
Table~\ref{Compare} summarizes and compares the above-reviewed physical-layer techniques for covert communication.
Generally, the existing covertness techniques can be classified into two categories in terms of the effects on adversaries. One is to mitigate the information signal leakage, 
and the other is to enlarge interference dynamics 
to cover the signal leakage. These approaches unavoidably consume additional system resources, such as bandwidth 
and energy, and sacrifice the communication performance at legitimate users.

 \begin{table*}  
 \caption {Comparison of existing  techniques for covert communication} \vspace{-7mm}
\begin{center}
\begin{tabular}{ p{25mm}   p{45mm} p{35mm}  p{35mm}  }  
\hline
Technique &    Effect on adversaries & Computational complexity & Deployment scalability \\  
\hline (Massive) Multiple antennas  &  Weakening signal leakage  & High & High   \\
\hline
AN generation  & Increasing interference dynamics  & Low if using a third-party device and high otherwise  &  Low if using a third-party device and high otherwise  \\
\hline
Cooperative relaying  & Weakening signal leakage & Low & Low \\
\hline
Spread spectrum  &  Weakening signal leakage  & High & High \\
\hline
MmWave &  Weakening signal leakage  & High & Low \\
\hline 
\end{tabular}
\end{center} \label{Compare} \vspace{-12mm}
\end{table*} 

\section*{IRS-Enhanced Covert Communication}

To tackle the resource-consuming issue of existing covertness techniques, we  
introduce an intelligent reflecting surface (IRS)-based solution to facilitate covert communication.  
The core technology of IRS is to have full control of electromagnetic behavior of the impinging waves by leveraging programmable metamaterials. Empowered by the IRS, the proposed approach has the potential to safeguard transmission from malicious detection 
by changing the propagation environment. 
{\em This approach is radically different from the existing ones since recycling the environment resources (i.e. transmitted signals that are not received by the intended receivers) has not been previously considered for covert communication}. An outstanding merit of the IRS that motivates the use of it for covert communication is its compatibility with the existing systems, to be illustrated in Fig. 2. In particular, an IRS can work in conjunction with existing covertness techniques without the necessity
to redesign the corresponding protocols and hardware, as an IRS only serves as an auxiliary device targeting on the manipulations of environmental signals.
Meanwhile, an IRS can be jointly configured with the existing system for performance optimization, an example of which is to be shown in the case study of this article.   
In the following, we first elaborate the basics of IRS, including the principles, features, and differences from other related concepts, and then introduce the IRS-enhanced covert communication systems.

\vspace{-2mm} 
\subsection*{Fundamentals of IRS}

An IRS is a software-controlled artificial surface that can be programmed to alter its electromagnetic response. 
The hardware realization of IRS is based on tunable metasurface, which is a thin and planar electromagnetic material 
consisting of 
discrete scattering particles spread over the structure,  
 the  
electromagnetic characteristics (e.g. capacitances and resonances) of which can be digitally re-engineered 
without re-fabrication. 
This can be realized by leveraging electronically tunable {\em meta-atoms},  such as liquid crystal, varactor/PIN
diodes, doped semiconductors,
micro-electro-mechanical systems (MEMS) switches, and flexible plasmonics. Generally, there exist three approaches to change the electromagnetic properties of meta-atoms, namely, {\em tunable resonator technique}, {\em guided-wave technique}, and {\em rotation technique},  a detailed review of which can be found in~\cite{V.Jan.2014Hum}. 

Configuring the constitutional meta-atoms collectively enables the entire metasurface to synthesize a wide diversity of radiation patterns that are infeasible with natural materials. 
The meta-atoms can either be tuned uniformly or individually. The former can realize simple electromagnetic manipulations such as absolute absorption and passive reflection, while the latter can support more complicated manipulations such as wave polarizing, imaging, and holograms. 
A distinguishing function of IRS enabled by the electromagnetic reconfiguration is to recycle existing environmental signals. Specifically, an IRS can reshape the phases, amplitudes, and reflecting angles of the environmental signals to serve its own objectives, e.g. jamming and signal cancellation.
With the striking advancement in fabrication techniques of metamaterials, modern IRSs are capable of fully reshaping the phase, amplitude, frequency, and reflecting
angles of impinging signals in a full-duplex fashion. 
For instance, the authors in~\cite{Kaina2014Nadge} implement a binary phase state IRS and demonstrate that for point-to-point transmission in an indoor environment the IRS can either boost the signal intensity  at the receiver by an order of magnitude or totally cancel it. More detailed knowledge of hardware fabrication and network implementation of IRS can be found in  \cite{V.Jan.2014Hum,D.May2019Renzo}.

Not to be confused with some  related techniques that can also be applied to facilitate covert communication, we discuss their key differences and highlight the comparative advantages of IRS as follows. 

\begin{itemize}

\item A {\em phased array} 
utilizes an array of radiators with variable phase shifts to create different beam patterns. As each radiator is associated with a dedicated active RF chain, a phased array incurs high hardware cost and appears with a large form factor. Moreover, the performance of a phased array degrades at high frequency (e.g. GHz)
as a result of reduced efficiency of the feed line. 
By contrast, 
an IRS features low-cost fabrications with nearly passive elements. Meta-materials such as ferroelectric films and graphene 
maintain good control of electromagnetic waves
over a wide frequency range covering Terahertz and visible region~\cite{V.Jan.2014Hum}. 
Another desirable feature of the IRS is that the contiguous surface  
enables more fine-grained spatial resolution of electromagnetic control than that of the spaced antenna arrays with radiator separation.

\item {\em Active metasurfaces}~\cite{S.May2018Hu}
make use of 
active materials (e.g. epsilon-near-zero materials~\cite{C.2013Jun}) to generate an electromagnetic field on the entire surface. Although active metasurface provides exceptional controllability of signals, the operation is energy-consuming and the configuration usually incurs high computational complexity, e.g. due to signal processing. 
By contrast, an IRS 
entails considerably reduced computational complexity and lower energy profile due to its passive electromagnetic manipulation.  
An IRS virtually consumes zero power during the reflecting process and incurs power consumption only when reconfiguring the electromagnetic properties of the IRS units.

\item {\em Full-duplex relays} resemble IRSs in the aspects of full-duplex transmission and multipath diversity gain.
Full-duplex relays can be either active or passive. 
\begin{itemize}
\item The active relay forwards data with its own signals generated from its active components (e.g. power amplifier). Due to full-duplex operation, the active relay inevitably causes self-interference and this signal processing latency. 
By contrast, an IRS is free from self-interference due to its passive electromagnetic operation.  

\item The passive relay reflects the existing source signals for data forwarding. The electromagnetic responses (i.e. reflection coefficients) of the passive relay are usually pre-designed and fixed. 
By contrast, an IRS possesses greater flexibility in adjusting its electromagnetic response.
\end{itemize}

Furthermore, an IRS is far more versatile than an information-forwarding relay as it can perform concurrent functions (e.g. beamsteering and interference cancellation) to satisfy heterogeneous quality-of-service (QoS) requirements.
 
\end{itemize}

Apart from 
its distinctive physical properties, 
IRSs are deployment-friendly.  
First, a metasurface can be fabricated with nearly passive elements (e.g. analog phase shifters) that do not rely on active components for transmission. Hence, the circuit power consumption of a metasurface is typically meager and can be powered through microwave energy harvesting. For example, the experimental results in~\cite{Kaina2014Nadge} show that the energy consumption of the implemented IRS is typically comparable to or lower than the amount of microwave energy it can recycle.
Second, thanks to the light-weight and ultra-thin footprint, metamaterials can be easily coated on the facade of environmental objects, e.g. walls, vehicles, and smart clothing, constructing a rich scattering environment.
Therefore, IRS-coated objects have the potential of  delivering a more deterministic wireless propagation environment in a self-sustainable manner. 
This opens up board opportunities to satisfy heterogeneous QoS requirements for future generation networks (e.g. more stable connectivity, improved data rate, higher spectral efficiency)  by only recycling existing environmental resources.

 \vspace{-2mm}
 \subsection*{IRS-Aided Covert Communication Systems}  

By leveraging the 
powerful electromagnetic control  
of metasurface, an IRS can be carefully designed to improve the undesirable propagation conditions to facilitate covert communication. Generally, there are two functions of IRS that can be utilized to enhance transmission covertness. On the one hand, an IRS can reflect the desirable signals (e.g. information transmission) in phase with the ones at the intended receiver so as to strengthen the signals,  referred to as {\em signal intensification}. 
On the other hand, an IRS can reflect the unwanted signals (e.g. information leakage and interference) in opposite phase with the ones at the unintended receiver, referred to as {\em signal cancellation}. Usually, there exists a trade-off in configuring the electromagnetic responses of the IRS elements to achieve the above two objectives simultaneously.

 \begin{figure}   \label{IRS}
 \centering
 \includegraphics[width=0.8\textwidth]{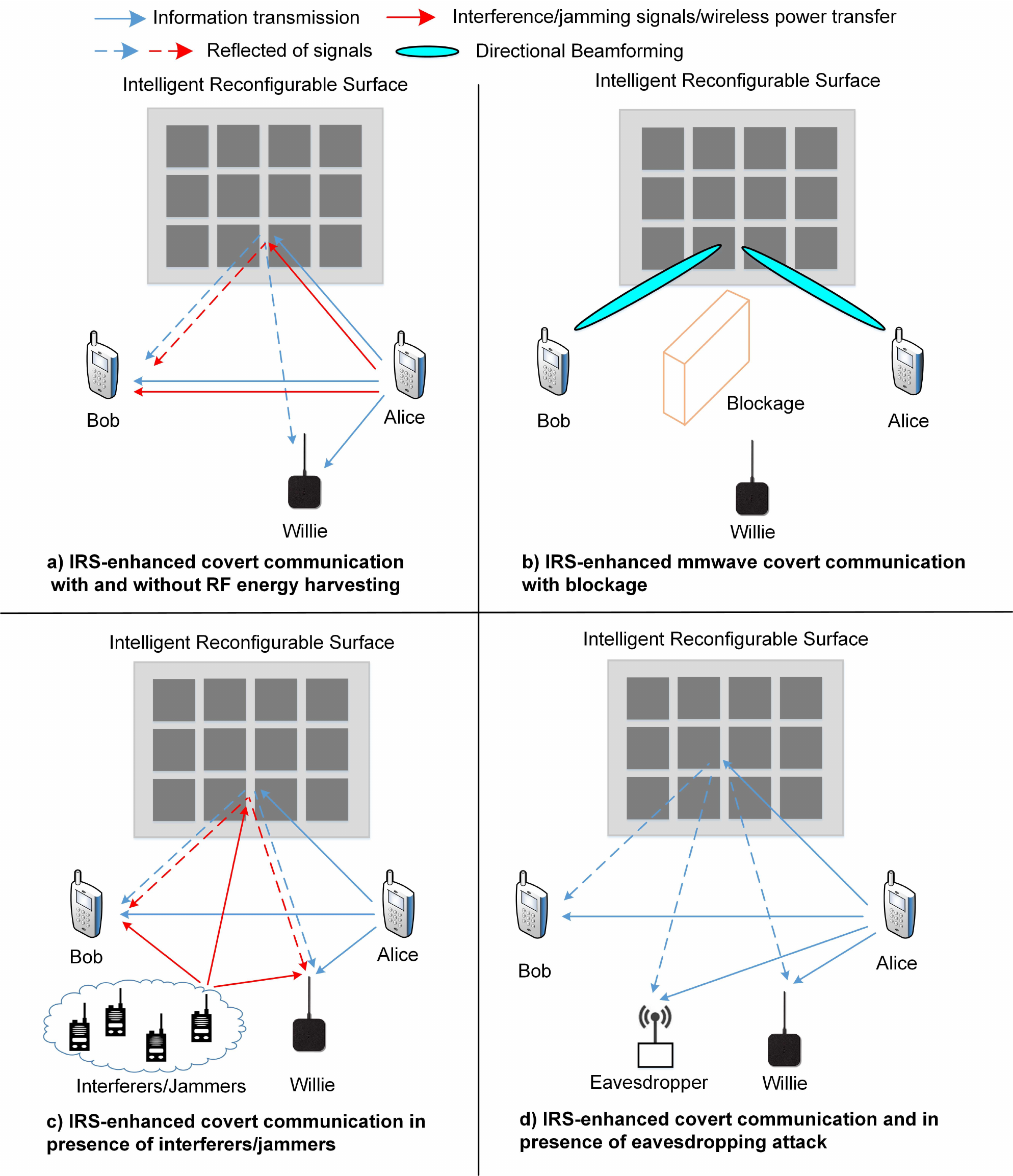}   
 \caption{IRS-enhanced covert communication systems.} \vspace{-8mm}
 \end{figure}

Next, we elucidate how an IRS can be exploited to enhance covert communication in various system environments (Fig. 2)\footnote{We note that Alice, Bob, Willie, jammer, eavesdropper and ambient transmitters illustrated in Fig. 2 can be any types of transceivers in practice.}.    
Fig. 2(a) illustrates a baseline system model where Alice intermittently transmits data to Bob in the presence of Willie. In this scenario, an IRS can be employed to perform signal intensification at Bob and signal cancellation at Willie. As a result, increased transmission rate at Bob and lower probability of detection at Willie could be achieved simultaneously. Additionally, in a wireless-powered covert communication system where Bob is equipped with RF energy harvesting capability, Alice can also perform wireless power transfer or simultaneous wireless information and power transfer to supply energy for Bob. Other than achieving covert communication, the IRS can also be utilized to facilitate wireless power transfer. 
Fig. 2(b) considers the scenario where Alice performs covert transmission with mmWave which is highly vulnerable to blockages due to severe penetration losses and poor diffraction of non line-of-sight (LoS) links. As shown, when there exists a  blockage between Alice and Bob, deploying an IRS with LoS links to both Alice and Bob can be used to address the negative impact of blockages for mmWave covert communication.
Fig. 2(c) depicts the scenario where the baseline system is affected by the co-channel interference, e.g., from ambient interferers or malicious jammers. In this case, the IRS can be additionally configured to conduct interference cancellation at Bob and interference intensification at Willie to conceal the signal from Alice.
In addition to the negative impact of co-channel interference, legitimate users may confront an eavesdropping attack as shown in Fig. 2(d).
Signal cancellation at both the eavesdropper and Willie needs to be conducted to cope with the concurrent attacks.

\vspace{-3mm}
\section*{Optimal Configuration for IRS-Enhanced Covert Communication under Noise Uncertainty: A Case Study}

We show a case study of designing an IRS-enhanced covert communication system. We consider noise as the only cover medium with the aim to focus on showing the effects of IRS on the covertness performance.
 
 \vspace{-3mm}
 \subsection*{System Model}
 \begin{figure}  [ht]
    \centering
     \vspace{-2mm} 
      {
    \label{fig:I_1}
   {
     \centering   
     \includegraphics[width=0.45 \textwidth]{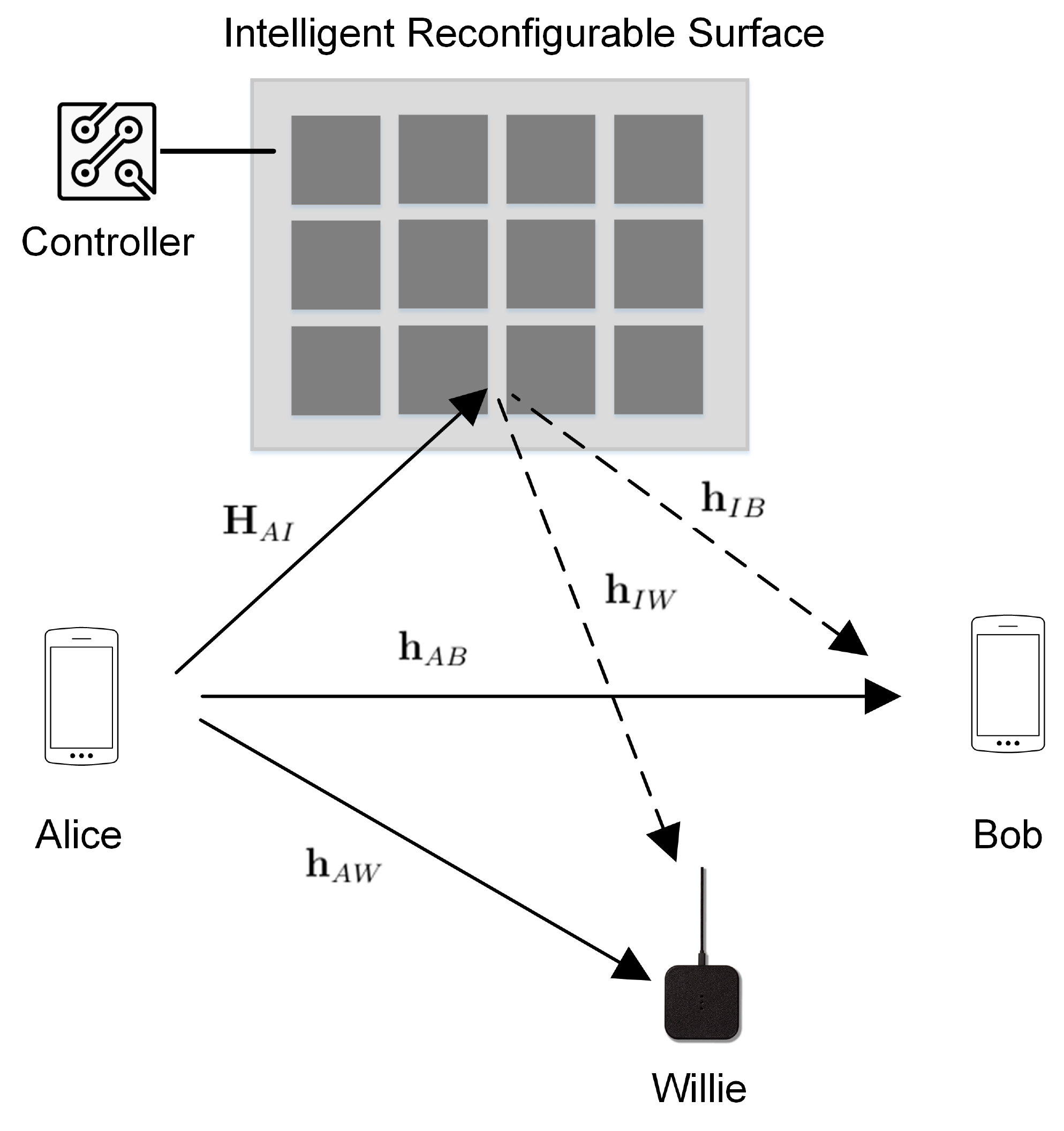}}}
     \centering 
    \caption{System model. 
    } 
    \centering
    \label{fig:tauAB}
    \vspace{-5mm}
    \end{figure}
    
We consider an IRS-enhanced covert communication system where Alice intends to transmit to Bob with LPD by a warden Willie.  
An IRS is deployed to facilitate the covert transmission of Alice. 
We consider that Alice, 
Bob, and Willie are all equipped with a single antenna. Alice has a maximum transmit power budget denoted by $\overline{P_{A}}$. 
The IRS consists of $N$ passive reflecting units, each of which can generate an arbitrary phase shift of the incident signal wave independently. All the channels in the system experience both power-law path loss with exponent $\alpha$ and block Rayleigh fading.   
Similar to \cite{M.CuiWCL}, we assume the CSI  
of all the channels are available at Alice and the IRS for the joint optimization, which yields the best system performance benchmark. It is worth noting that the CSI of Willie can be reasonably estimated when Willie is an active transmitter \cite{A.2014Mukherjee}. 
 
We consider a bounded uncertainty model for the noise observed by Willie  $\sigma^2_{W}$, the PDF of which is given by \cite[eqn. 3]{B.April2017He}: $f_{\sigma^2_{W}}(x)= 1/ \big( 2 \ln (\rho) x\big)$,  if $\sigma^2_{n}/ \rho \leq \sigma^2_{W} \leq  \sigma^2_{n} \rho$ and 
$f_{\sigma^2_{W}}(x)=0$, otherwise. Here, $\sigma^2_{n}$ is the nominal noise
power, $\rho \in [1, \infty)$ is the uncertainty parameter which determines the range of $\sigma^2_{W}$. 
Note that a larger value of $\rho$ represents larger uncertainty of $\sigma^2_{W}$. Similar to \cite{B.April2017He}, the noise uncertainty at Bob is not considered as it does not affect the  covertness performance. 
Specifically, the noise power at Willie is considered as Gaussian white noise with zero mean and variance $\sigma^2_{B}$. Moreover, Willie is considered to know a priori distributions of $\sigma^2_{W}$ and the received signals from Alice, however, is unaware of the existence and operation of the IRS. Thus, Willie can only set its detection threshold based on the available a priori knowledge.   
 
We consider the optimization problem to maximize the covert rate between Alice and Bob measured by the Shannon capacity through jointly optimizing the phases of the IRS units and the transmit power of Alice subject to 1) the {\em covertness constraint} that the sum of the false alarm probability and mis-detection probability of Willie is greater than a target threshold $\xi$; 2) the {\em phase operation constraint} that the phase shifts of the IRS units are within $[0, 2 \pi )$; 3) the {\em transmit power constraint} that Alice should transmit at a power level below $\overline{P}$.

\vspace{-5mm}
\subsection*{Numerical Results}
 \begin{figure}  
 \centering
  \begin{minipage}[c]{0.48\textwidth}
  \includegraphics[width=0.95\textwidth]{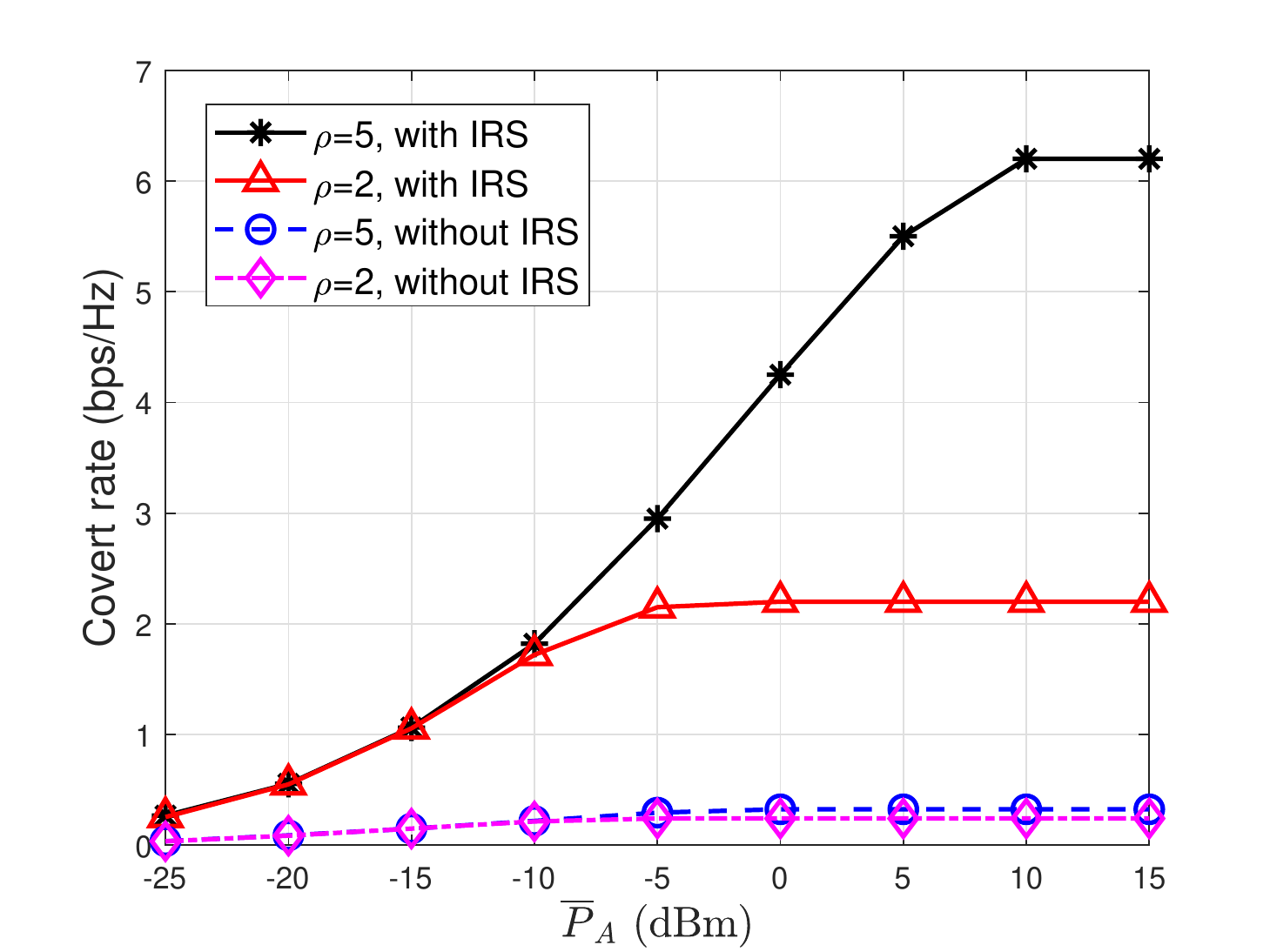} \vspace{-2mm}
  \caption{Covert rate as a function of $\overline{P}_{A}$ ($N=25$, $d=10$, $\sigma^2_{n}=\sigma^2_{B}=-60$ dBm, $\xi=99\%$, $\alpha=3$). 
  } \label{fig:Rate_PA}
  \end{minipage} 
   \begin{minipage}[c]{0.48\textwidth}
   \includegraphics[width=0.95\textwidth]{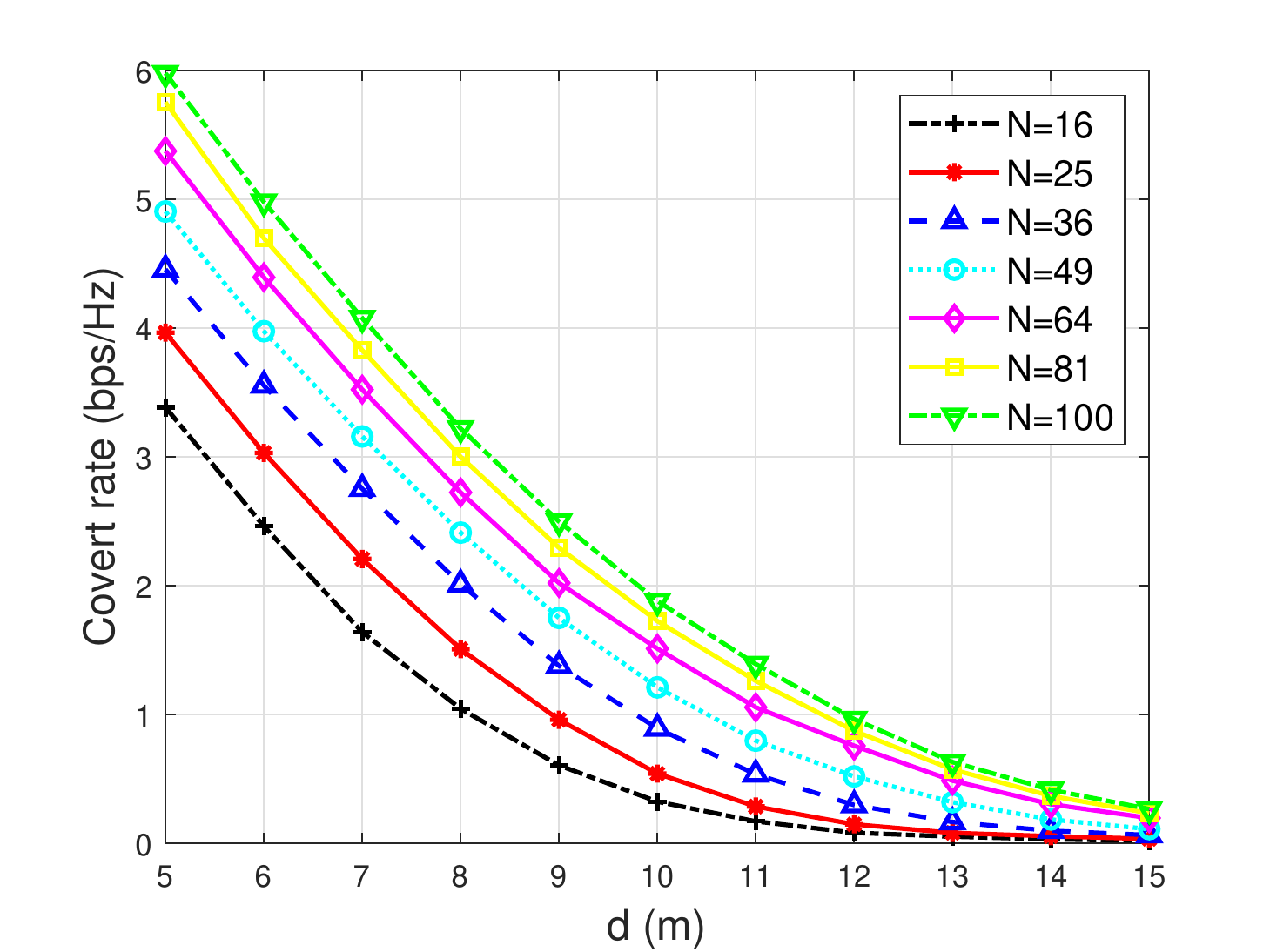}  \vspace{-2mm}
   \caption{Covert rate as a function of $d$ ($\overline{P}$, $\rho=5$, $\sigma^2_{n}=\sigma^2_{B}=-30$ dBm, $\xi=99\%$,  $\alpha=3$). }
   \label{fig:Rate_D} 
   \end{minipage}
  \vspace{-5mm} 
  \end{figure}
 
Next, we perform Monte Carlo simulations to study the formulated optimization problem. In the simulations, Alice, Bob, the IRS, and Willie are located at (0, 0), ($d$, 0), ($d/2$, 0),
and (0, 15)  in a two-dimensional area, respectively. 
Alice is considered to transmit with a probability of 50$\%$. For each simulation realization, we generate independent and identically distributed noise power at Willie $\sigma^2_{W}$ and fading gains for all the channels  which are exponentially distributed with unit mean.
 
Fig. \ref{fig:Rate_PA} depicts the covert rate $R_{AB}$ as a function of the maximum transmit power $\overline{P}_{A}$. For the comparison purpose, we also present the results of $R_{AB}$ without the use of an IRS. It can be found that $R_{AB}$ can be dramatically improved with the aid of an IRS. Moreover, $R_{AB}$ reaches a steady value at much larger  $\overline{P}_{A}$ in the case with an IRS compared to that without an IRS. The reason that $R_{AB}$ stops increasing at a steady value is that, given a certain noise level, the covertness constraint can no longer be maintained with the help of the IRS if the transmit power of Alice is above a certain level. 
Another observation is that greater noise uncertainty at Willie (represented by larger $\rho$) helps to improve $R_{AB}$. The performance gap between the cases with $\rho=5$ and $\rho=2$ becomes much more conspicuous with the increase of $\overline{P}_{A}$.
 
Fig.~\ref{fig:Rate_D} demonstrates 
the impact of the number of IRS elements $N$ under varying transmission distance.
The results show that larger $N$ renders better performance, especially when $d$ is large.  For instance, the ratio of $R_{AB}$ with $N=64$ to that with $N=16$ is $163.6\%$ when $d=5$ and is increased to  $523.2\%$  when $d=10$. This also implies that employing more IRS elements is an effective way to improve the covert transmission distance. For increase, if the target covert rate is 1 bps/Hz, the covert transmission distance is extended  from about 8 m to about 11.7 m when $N$ is increased from $16$ to $64$.

\vspace{-3mm}

\section*{Concluding Remarks and Future Directions}

With the integration of the IRS to covert communication systems, the previously unused environment resources can be recycled to enhance communication covertness. 
The article reviews the existing covertness techniques and envisions the use of the IRS to revolutionize covert communication systems in various aspects. A case study has also been presented to demonstrate that a considerable improvement of covert performance can be achieved through 
the joint configuration of the IRS and the covert communication system.  We firmly believe that the emerging IRS technology will open up broad  opportunities in designing and developing future wireless security, not limited to 
 covertness techniques.   

The scope of future research topics on IRS-enhanced covert communication is broad. Some open issues and research directions are as follows:

\begin{itemize}
\item  {\em Estimation of IRS channels}: 
As the wave manipulation of an IRS is dependent on the CSI to enhance covert communication, the system performance is heavily dependent on the availability and accuracy of CSI. However, instantaneous CSI of the reflection channels is difficult to be acquired due to the nearly passive operation of an IRS. In this context, the machine learning-based approach that allows estimating channels without explicit feedback/detection is worth exploring to devise feasible solutions. 

\item {\em IRS-based information/communication theoretic models}: 
With the signal intensification and cancellation capabilities of the IRS, an IRS-enhanced covert channel is expected to transport a larger volume of information bits.
Hence, the conventional covert channel capacity needs to be revisited by taking into account channel programmability. 
Moreover, scaling laws of IRS-enhanced covert channel capacity needs to be derived for a fundamental understanding of achievable performance limits.

\item  {\em Impact of multiple IRSs}: 
IRSs are anticipated to be deployed on the superficies of environmental objects located with perplexing spatial patterns. Therefore, it is a common scenario that the propagation environment is jointly shaped by multiple IRSs. The aggregated impact of the operation of ambient IRSs on IRS-enhanced covert communication is worth investigating by considering their spatial distribution. 

\end{itemize}


\vspace{-2mm}
 \begin{thebibliography}{99}
\bibitem{I.2018Ahmad}
I. Ahmad, T. Kumar, M. Liyanage, J. Okwuibe, M. Ylianttila, and A. Gurtov, ``Overview of 5G security challenges and solutions,"
\emph{IEEE Communications Standards Magazine}, vol. 2, no. 1, March 2018. 

 \bibitem{Y.2009Liang}
 Y. Liang, H. V. Poor, and S. Shamai, ``Information theoretic security," \emph{Foundations and Trends in Communications and Information Theory}, vol. 5, pp. 355-580, June 2009. 
 
\bibitem{A.Sep.2013Bash} 
B. A. Bash, D. Goeckel, and D. Towsley, ``Limits of reliable communication with low probability of detection on AWGN channels," \emph{IEEE J. Sel. Areas Commun.}, vol. 31, no. 9, pp. 1921-1930, Sep. 2013. 
 

  
 \bibitem{M.CuiWCL}
 M. Cui, G. Zhang, and R. Zhang, ``Secure wireless communication via intelligent reflecting surface," \emph{IEEE Wireless Communications Letters}, to appear.


\bibitem{X.2019Zheng}
T. X. Zheng, H. M. Wang, D. W. K. Ng, and J. Yuan,
``Multi-antenna covert communications in random wireless networks," 
\emph{IEEE Transactions on Wireless Communications}, vol. 18, no. 3, March  2019.
 
 
 
 \bibitem{E.Bjornson}
E. Bjornson, {\em et al}., “Massive MIMO systems with non-ideal hardware: Energy efficiency, estimation, and capacity limits,” \emph{IEEE Transactions on Information Theory}, vol. 60, no. 11, Nov.  2014. 
 
 \bibitem{R.2018Soltani}
R. Soltani, \emph{et al.}, ``Covert wireless communication with artificial noise generation," \emph{IEEE Transactions on Wireless Communications}, vol. 17, no. 11,  pp. 7252-7267, Nov. 2018.
 
\bibitem{G.2014Andrews}
J. G. Andrews, {\em et al}, ``What will 5G be?," \emph{IEEE J. Sel. Areas Commun.} vol. 32, no. 6, pp.  1065-1082, June  2014. 


\bibitem{V.Jan.2014Hum}
S. V. Hum and J. P. Carrier, 
``Reconfigurable reflectarrays and array lenses for
dynamic antenna beam control: A review," \emph{IEEE Transactions on Antennas and Propagation}, vol. 62, no. 1, Jan. 2014.
 
\bibitem{Kaina2014Nadge}
N. Kaina, {\em et al.}, ``Shaping complex microwave fields in reverberating media with binary tunable metasurfaces,” Scientific reports, 4, 6693.
Oct. 2014.


\bibitem{D.May2019Renzo}
M. D. Renzo, {et al.,}, ``Smart radio environments empowered by reconfigurable AI meta-surfaces: an idea whose time has come," \emph{EURASIP Journal on Wireless Communications and Networking},  vol. 1, pp. 1-20, May 2019.

\bibitem{S.May2018Hu}
S. Hu, F. Rusek, and O. Edfors, ``Beyond massive MIMO: The potential of data transmission with large intelligent surfaces," \emph{IEEE Trans. Signal
Process.}, vol. 66 , no. 10, May 2018.


\bibitem{C.2013Jun}
Y. C. Jun, J. Reno, T. Ribaudo {\em et al.}, ``Epsilon-near-zero strong
coupling in metamaterial-semiconductor hybrid structures," \emph{Nano Letters}, vol. 13, no. 11, pp. 5391–5396, Oct. 2013.
 
 


\bibitem{A.2014Mukherjee}
A. Mukherjee, S. A. A. Fakoorian, J. Huang, and A. L. Swindlehurst, ``Principles of physical layer security in multiuser wireless networks: A
survey," \emph{IEEE Commun. Surveys Tuts.}, vol. 16, no. 3, pp. 1550-1573, Third Quarter 2014.


\bibitem{B.April2017He}
B. He, {\em et al}, ``On covert communication with noise uncertainty," \emph{IEEE Communications Letters}, vol. 21, no. 4,  April 2017.

\end{thebibliography}
\end{document}